\documentclass{aa}  

\usepackage{graphicx}
\usepackage[dvipsnames]{xcolor}
\usepackage{txfonts}
\usepackage{booktabs,multirow,array,caption}
\usepackage[colorlinks=true,
    linkcolor=blue,
    citecolor=blue,
    filecolor=magenta,      
    urlcolor=cyan]{hyperref}
\usepackage{ulem}
\usepackage{orcidlink}

\begin{document}

   \title{An effective model for the tidal disruption of
satellites undergoing minor mergers with axisymmetric primaries}
   \titlerunning{Tidal disruption of satellite galaxies}
   \authorrunning{Varisco et al.}

   \subtitle{}
 \author{Ludovica Varisco\orcidlink{0000-0002-6724-5999}
          \inst{1,}
          \inst{2} 
          \fnmsep\thanks{l.varisco4@campus.unimib.it}
        \and
        Massimo Dotti
          \inst{1,}
          \inst{2,} 
          \inst{3}
          \and
          Matteo Bonetti\orcidlink{0000-0001-7889-6810}
          \inst{1,}
          \inst{2,}
          \inst{3}
          \and
          Elisa Bortolas
          \inst{1,}
          \inst{2}        
          \and
          Alessandro Lupi
          \inst{4,}
          \inst{2,}
          \inst{1}
    }

   \institute{
            Universit\`a degli Studi di Milano-Bicocca, Piazza della Scienza 3, I-20126 Milano, Italy
        \and
            INFN, Sezione di Milano-Bicocca, Piazza della Scienza 3, I-20126 Milano, Italy
        \and
            INAF - Osservatorio Astronomico di Brera, via Brera 20, I-20121 Milano, Italy
        \and
            DiSAT, Università degli Studi dell’Insubria, via Valleggio 11, I-22100 Como, Italy
    }

   \date{}

% \abstract{}{}{}{}{} 
% 5 {} token are mandatory
 
  \abstract{
   According to the hierarchical formation paradigm, galaxies form through mergers of smaller entities and massive black holes (MBHs), if lurking at their centers, migrate to the nucleus of the newly formed galaxy, where they form binary systems. The formation and evolution of MBH binaries, and in particular their coalescence timescale, is very relevant for current and future facilities aimed at detecting the gravitational-wave signal produced by the MBH close to coalescence. While most of the studies targeting this process are based on hydrodynamic simulations, the high computational cost makes a complete parameter space exploration prohibitive. Semi-analytic approaches represent a valid alternative, but they require ad-hoc prescriptions for the mass loss of the merging galaxies in minor mergers due to tidal stripping, which is not commonly considered or at most modelled assuming very idealised geometries. In this work, we propose a novel, effective model for the tidal stripping in axisymmetric potentials, to be implemented in semi-analytic models. We validate our semi-analytic approach against N-body simulations considering different galaxy sizes, inclinations, and eccentricities, finding only a moderate dependence on the orbit eccentricity. In particular, we find that, for almost circular orbits, our model mildly overestimates the mass loss, and this is due to the adjustment of the stellar distribution after the mass is removed. Nonetheless, the model exhibits a very good agreement with simulations in all the considered conditions, and thus represents an extremely powerful addition to semi-analytic calculations.}


   \maketitle
%
%-------------------------------------------------------------------

\section{Introduction}
In the framework of the hierarchical paradigm of cosmic structure formation \citep{1974ApJ...187..425P,1978MNRAS.183..341W}, galaxies form in a bottom-up fashion, whereby the massive galaxies that we see today build up at the intersection of dark matter filaments along which other galaxies and cold gas can stream inwards \citep{2009ApJ...703..785D}. Specifically, at those ``cosmic crossroads'', galaxies are expected to experience a sequence of mergers and accretion events that contribute to their final mass and morphological appearance. 

Galactic mergers are categorised based on the mass-ratio of the involved galaxy pairs.
The threshold between minor and major mergers is not  universally determined as different values are employed in literature depending on the specific research objectives and contexts. \cite{2011ApJ...729...85C} classify as major mergers systems involving galaxies with mass ratio exceeding 1:10, while those falling below this value are designated as minor mergers. A widely used classification defines as major mergers systems with mass ratio grater than 1:3, while minor merger as those falling in the range 1:3-1:10 \citep[e.g.][]{Cox_2008,2009MNRAS.397..802H, 2021MNRAS.503.3113M, 2009ApJ...706...67R}. \cite{2019A&A...631A..87V} categorises galaxy pairs on the basis of the stellar mass ratio of the systems involved, defining as major, minor, and very minor mergers the systems corresponding to the ranges 1:1-1:6, 1:6-1:100 and $<$ 1:100, respectively.

Besides the specific choice of the threshold, the distinction between minor and major mergers is not a mere classification, but implies very different dynamical evolution, outcomes and investigation techniques.
Major mergers are generally rare and disruptive events that completely reshuffle the material in the parent systems and significantly perturb the original morphology in a few dynamical times. Given their disruptive effect, they can be properly characterised only through expensive numerical simulations able to track the strongly time-varying gravitational potential. 
On the contrary, minor mergers are usually common events along galaxy lifetimes and they generally represent a (small to moderate) perturbation to the more massive system in which they sink. In this regard, the secondary galaxies involved in minor galaxy mergers can be treated as massive perturbers, i.e. objects considerably heavier than the single bodies forming the galactic structure, but much less massive than the whole host galaxy. By leaving the more massive galaxy nearly unchanged, minor mergers are suitable to be modelled in a semi-analytical fashion \citep[e.g.][]{2012MNRAS.425.3119H}. 
This feature opens the possibility of performing investigations with inexpensive computational loads, still requiring a proper and careful tuning of the semi-analytical recipes against numerical simulations.
\vspace{3cm}

Even though single minor mergers do not typically produce morphological transformations of host galaxies \footnote{See however \cite{Jackson_2019}, who demonstrate that single minor merger events involving systems with mass ratios $\sim$ 0.1-0.3, and with the satellite moving on orbits almost aligned with the host's disc plane, may trigger catastrophic changes in the primary morphology within timescales as short as a few hundreds Myr.}, recent theoretical and observational studies highlight the important role that repeated minor mergers may play on the evolution of their massive companions. Indeed, the occurrence of multiple minor mergers in disc galaxies can gradually induce a significant redistribution of the stellar orbits in the primary system, thus forming slowly rotating spheroidal remnants \citep[see e.g.][]{2007A&A...476.1179B,2011A&A...535A...5Q, 2013ApJ...778...61T}. \cite{Martin_2018} showed that one third of the morphological transformation of galaxies undergoing galaxy mergers over the cosmic time is due to repeated minor merger events, the latter becoming the dominant driver of morphological changes at late epochs ($z\gtrsim 1$). 
Moreover, minor mergers have been proven to enhance both star formation, being responsible for over a half of the star formation events induced by galaxy mergers in the Universe, and massive black holes (MBHs) accretion rates \citep{10.1093/mnras/stu338,Pace_2014}, and also to be responsible for the $70\%$ of the merger-driven asymmetric structures in post-merger galaxy remnants \citep{2024MNRAS.527.6506B}.

Among massive perturbers inhabiting galaxies, MBHs are particularly interesting to study. MBHs are located in the nuclei of most of massive galaxies (if not all of them) \citep[see e.g.][]{2005SSRv..116..523F,2013ARA&A..51..511K} and through galaxy mergers multiple MBHs are delivered within the same host, eventually leading to the formation of massive black hole binaries (MBHB), triplets or even higher order multiplets \citep{1980Natur.287..307B,2003ApJ...582..559V,2018MNRAS.477.2599B}. These systems are primary targets of current and forthcoming gravitational wave (GW) experiments, primarily Pulsar Timing Array \citep[PTA,][]{2023ApJ...951L...8A,2023arXiv230616214A,2023ApJ...951L...6R} campaigns now opening the nHz sky, and the Laser Interferometer Space Antenna (LISA), targeting mHz frequencies \citep{2017arXiv170200786A}.
Prior to the formation of bound MBH systems in the nuclei of galaxies, every MBH needs to sink towards the central regions. The main actor driving this evolution is  dynamical friction \citep[DF,][]{1943ApJ....97..255C}.
At this stage of the evolution MBHs are generally still surrounded by their progenitors' cores, so that their effective sinking mass (locally perturbing the primary and leading to DF) can be much larger than the mass of the MBH itself \citep{2020A&ARv..28....4N}. However, such left-over material (gas and stars) surrounding the MBH typically gets gradually stripped by the main galaxy tidal field \citep{1987gady.book.....B}. The effectiveness of the process depends on the compactness of the material around the intruder MBH, and on the steepness of the galactic acceleration field. Depending on the efficiency of the stripping process, the MBH loses material and may eventually ``get naked'', i.e. remain without any residual surrounding distribution of matter bound to it. This effective ``mass loss'' crucially affects the dynamics of the inspiral and especially the efficiency of DF, as the DF timescale needed for the object to reach the centre of the primary galaxy critically depends on the perturber's mass \citep[see e.g.][]{2020MNRAS.498.2219V}.

A quantitative assessment of how mass is stripped from infalling satellite galaxies requires a careful estimation of the so called tidal radius, i.e. the conceptual boundary for a celestial object dividing the bound from the unbound mass. Beyond this limit, the object's material undergoes stripping due to the tidal field of the more massive companion. First introduced by \citet{1957ApJ...125..451V} within the context of Milky Way globular clusters, the tidal radius is theoretically defined strictly for satellites following circular orbits, where it coincides with the position of L1/L2 Lagrange points \citep{1987gady.book.....B}. A different attempt to define such radius also for eccentric motion was explored by \citet{1962AJ.....67..471K}, who argued that during pericenter passages, satellites are truncated to the size indicated by the pericentric tidal radius.
Later, \citet{1970A&A.....9...24H} and \citet{1975CeMec..11...85K} observed that retrograde orbits in the context of the restricted three-body problem are stable over greater distances compared to prograde orbits, further out the tidal radius defined by \citet{1962AJ.....67..471K}. In a more recent study, \citet{2006MNRAS.366..429R} derived an expression for the tidal radius taking into different orbit types: prograde, radial, and retrograde. Interestingly, the analysis revealed that the tidal radius for retrograde orbits exceeds that of radial orbits, which, in turn, is larger than the tidal radius for prograde orbits.

To date, the vast majority of attempts to estimate the tidal radius focused on spherically symmetric host galaxies \citep[see however][]{2016ApJ...819...20G}. 
Although observations show that, while the morphology of massive galaxies in local Universe is dominated by spheroidal systems \citep[see e.g.][]{2003AJ....125.1817B,2014MNRAS.444.1125C}, in the early Universe the massive galaxy population was mostly composed of disc galaxies \citep[see e.g.][]{2014MNRAS.439.1494B,2015ApJS..219...15S}. This morphological transformation which leads to an overall transition from rotationally-supported systems to dispersion dominated ones is believed to be primarily driven by galaxy mergers. Moreover, cosmological simulations suggest that disc galaxies do not show any significant difference in their merger history compared to spheroidal galaxies \citep[see e.g.][]{Martin_2018}. Thus, a significant number of mergers involving disc-like primary galaxies are expected to have occurred throughout cosmic history and are still ongoing. Indeed, observations on nearby massive disc galaxies display tidal features, hinting that they have undergone recent minor mergers events. For this reason, a systematic investigation focused on galaxy mergers involving systems that strongly deviate from spherical symmetry is compelling.

In this study, we precisely aim at finding a general description of the tidal radius when axis-symmetric systems are involved\footnote{Here, we refer to the total potential of the primary galaxy, composed of both baryonic and dark matter components. If one focuses on the dark matter halo, the work of \cite{Kazantzidis_2004} show that baryonic cooling and the formation of a disk can enhance symmetry in the inner regions of halos. }. Those systems, representative of e.g., spiral galaxies, are indeed quite common and many minor mergers actually occur in such  galaxies. Our ultimate goal consists in deriving a simplified prescription for the tidal radius to be implemented in semi-analytical models of galaxy formation, in order to better asses the DF-driven inspiral pace of massive perturbers within galaxies of any type.

A proper and comprehensive semi-analytical modellisation of minor mergers can represent a powerful tool for studying a wide variety of astrophysical scenarios. The exploitation of semi-analytical models is crucial to overcome the limited spacial and mass resolution of large-scale cosmological simulations. In these simulations, numerous minor mergers are observed to occur, however the lack of sufficient resolution may hinder to track the late stages of these events as the satellite galaxies become unresolved. Employing detailed semi-analytical models would enable us to follow the satellite evolution down to scales where the system is no longer resolved in the simulations. This feature allows us to predict the late phases of the merger and to determine the ultimate fate of the satellite galaxy and, if present, of the MBH embedded within it. In this context, semi-analytical models could be useful, for instance, to address and possibly reconcile discrepancies between the estimated fraction of orphan galaxies arising from mock and semi-empirical models \citep[see e.g.][]{kumar2023overabundance}.
Furthermore, due to their great versatility, semi-analytical models are particularly well-suited for studying the formation and evolution of systems in extreme merger scenarios, such as very faint Milky Way satellites \citep{smith2023discovery}. Finally, minor mergers may also trigger an enhancement in the satellite MBH accretion due to gas inflows caused either by shocks developing within the interstellar medium in the pairing phase at the contact surface of the two galaxies \citep{2017MNRAS.465.2643C}, or in the final phases when the naked MBH circularises inside the primary disk \citep{2011ApJ...729...85C}. 

The paper is organised as follows: in Sec. \ref{sec:methods}, we introduce a novel prescription for the tidal radius, delineate the galactic models employed, and detail the setup of the N-body simulations implemented for our prescription validation. In Sec. \ref{sec:results}, we present the outcomes of the comparison between out model's predictions and those derived from N-body simulations. Finally, in Sec. \ref{sec:conclusions} we discuss the limitations of our model, we summarise our findings and draws our conclusions.

\section{Methods}
\label{sec:methods}

When minor mergers occur, satellite galaxies, while orbiting within their hosts, are subjected to tidal forces that remove part of their mass, sometimes leading to their complete disruption even after a single pericentre passage. Two main mechanisms have been identified for removing mass from the satellite, depending on the rapidity at which the external tidal field varies. When the satellite experiences a slowly changing tidal field, the effect of the tidal forces is that of stripping material from the outer regions of the satellite, forming a clear external boundary often called the tidal radius ($R_t$). 
This process is identified as \textit{tidal stripping}. On the contrary, when the satellite undergoes a rapid change in the external tidal field, part of its orbital energy is converted into internal energy, leading to an overall heating of the satellite. The amount of energy injected into the system during fast pericentre passages and transferred to the stars can be enough to unbind a significant fraction of the satellite mass. This effect is known as  \textit{tidal heating}. 

The mass loss caused by tidal effects can significantly impact the orbital decay of the satellite, as it reduces the efficiency at which dynamical friction drags the satellite galaxy towards the centre of its host, thus increasing its orbital decay time.

\subsection{Tidal Radius} \label{sec:Tid_rad}
To characterise the mass loss of satellite galaxies due to tidal stripping in minor mergers, the first step consists of defining the tidal radius. The standard approach in literature considers two spherically symmetric systems, with mass profiles $m(r)$ for the satellite, and $M(r)$ for the host galaxy, whose centres are separated by a distance $R$. The satellite $R_t$ is defined as the distance from the centre of the satellite at which the acceleration of a test particle along the direction connecting the centre of the two systems vanishes. In a minor merger scenario where $m \ll M$, under the assumptions that $R_t \ll R$ at any time, and that the test particle has null velocity in the satellite's reference frame, $R_t$ is given by: 
\begin{equation}
    R_t = R \bigg [ \frac{G m(R_t)}{\Omega^2- \frac{\mathrm{d^2} \Phi_h}{\mathrm{d}r^2}}  \bigg]^{\frac{1}{3}}.
\label{Eq:TR_king}
\end{equation}

This expression was first  derived in \cite{1962AJ.....67..471K}, where $r$ and $\Omega$ are the radial coordinate and  the angular velocity of the satellite in the reference frame of the host galaxy, and $\Phi_h (r)$ is its gravitational potential.
It is worth noting that this formula is strictly valid  for circular orbits, but can be easily extended to eccentric orbits if one considers instantaneous values for $\Omega$ and $R$. Additionally, it is important to emphasise that Eq.~\eqref{Eq:TR_king} holds only under the simplistic assumption of a spherical host. 

In this study, we aim to present a novel prescription for $R_t$ that is adaptable to various host geometries. For this purpose, we consider a spherically symmetric satellite galaxy embedded in the generic potential of its host.
We define the galactic inertial frame with the origin in the galactic centre denoted as $S$ and the non-inertial frame of the satellite as $S'$. In this work, all the quantities evaluated in the non-inertial frame of the satellite are primed, while the unprimed are relative to the inertial frame of the host galaxy.
 Considering a test satellite star, its position is identified by the radius vector $\mathbf{r_*}$.
The acceleration of the test star in the reference frame of the satellite is:
\begin{equation}
    \mathbf{a'} = \mathbf{a} - \mathbf{A} - \frac{d \mathbf{\Omega}}{d t} \times \mathbf{r'_*} - \mathbf{\Omega} \times (\mathbf{\Omega} \times \mathbf{r'_*}) - 2\mathbf{\Omega} \times \mathbf{v'}.
\end{equation}
Here, $\Omega$ is the angular velocity of the satellite centre of mass (CoM), $\mathbf{a}$ represents the acceleration of the test star in the $S$ frame:
\begin{equation}
    \mathbf{a} = - \frac{GM_s(r'_*)}{r'^{3}_*} \mathbf{r'_*} - \mathbf{\nabla} \phi_h(r_*),
\end{equation}
and $\mathbf{A}$ is the acceleration of the $S'$ frame in  $S$, which can be expressed as:
\begin{equation}
    \mathbf{A} = - \mathbf{\nabla} \phi_h(r_S),
\end{equation}
where $r_S$ indicates the distance of the satellite CoM from the host's centre.
The term $\mathbf{\Omega} \times (\mathbf{\Omega} \times \mathbf{r'_*})$ can be rewritten as $\Omega^2 r'_*(\cos \alpha-1)$, with $\alpha$ being the angle between $\mathbf{\Omega}$ and $\mathbf{r'_*}$.
Choosing a random direction $\mathbf{\hat{e}_{r'*}}$ from the centre of the satellite, we can approximate the tidal radius as the distance from the satellite centre at which a test star with $v' = 0$ experiences a vanishing $\mathbf{a'}$:
\begin{equation}
    a'_{\mathbf{\hat{e}_{r'_*} }}= - \frac{G M_s(r'_*)}{r'^2_*}  - \mathbf{\nabla} \phi_h(\mathbf{r_*})  \cdot  \mathbf{\hat{e}_{r'*}} + \mathbf{\nabla} \phi_h(\mathbf{r_S})  \cdot  \mathbf{\hat{e}_{r'*}} - \Omega^2 r'_*(\cos \alpha-1),
\label{Eq:new_def_TR}
\end{equation}
where we omitted the term  $d\Omega/ d t \times \mathbf{r'_*}$ which is directed perpendicularly to $\mathbf{\hat{e}_{r'*}}$, thus not contributing to the acceleration along the reference direction we fixed. It is important to note that, unlike the derivation in \cite{1962AJ.....67..471K}, we relax the assumption $R_t \ll R$, therefore allowing the satellite to undergo close encounters with the host centre.
Eq.~\eqref{Eq:new_def_TR} thus provides an implicit definition for $R_t$ along a specific direction from the centre of the satellite. As mentioned above, if the host system is  spherically symmetric, the reference direction along which the $R_t$ is evaluated is the one connecting the centre of the two galaxies, since it is the direction that maximises the tidal force. However, in a generic galactic field it is not possible a priori to define the direction that maximises the tidal force exerted on the satellite by the host, which instead will depend on the morphological parameters of the two systems and the instantaneous location of the satellite within the host potential. For this reason, at any time during the satellite evolution we numerically solve Eq.~\eqref{Eq:new_def_TR} along 1000 random directions and we select $R_t$ as the minimum of all the tidal radii evaluated, that we denote as $R_{T1}$.
However, the mass of the satellite is not instantaneously stripped and it is not possible a priori to define at which rate the material is removed through Eq.~\eqref{Eq:new_def_TR}. For this reason, we introduce a modified definition of the tidal radius, i.e.
\begin{equation}
    R_{T2}(t) = R_T(t_{\rm{old}})\;  e^{{-\alpha \frac{t - t_{\rm old}}{r_p / v_p}}}.
\label{eq:TR_2}
\end{equation}
In Eq.~\eqref{eq:TR_2}, $ R_T(t_{\rm{old}})$ is the tidal radius evaluated at a prior time $t_{\rm old}$, $r_p$ and $v_p$ are the distance and velocity of the satellite with respect to the host centre both evaluated at the pericentre, while $\alpha$ is a tunable dimensionless parameter that regulates the rate at which the mass is removed from the satellite: the higher the value of $\alpha$, the faster the mass is stripped.
Thus, comparing $R_{T1}$ and $R_{T2}$ we define $R_{T}$ to be:
\begin{equation}
    R_T(t) = \max(R_{T1}(t), R_{T2}(t)).
\end{equation}

Finally, we require the tidal radius to be a decreasing function of time. This condition implies that the removed material is irrevocably detached from the satellite, precluding any subsequent reattachment in later times, effectively assuming that tidal stripping is irreversible.

\subsection{Satellite galaxy}
In this study, we characterise the satellite galaxy employing the spherical and isotropic Hernquist model \citep{1990ApJ...356..359H}, whose potential and associated mass density profile are given by:
\begin{equation}
    \Phi_s(r) = - \frac{G M_s}{r+a_s},
\end{equation}
\begin{equation}
   \rho_s (r) = \frac{M_s}{2 \pi} \frac{a_s}{r(r+a_s)^3},
\label{eq:hern_dens}
\end{equation}
where $M_s$ and $a_s$ are the total mass and scale radius of the satellite, respectively. The corresponding mass profile is $ m_s (r) = M_s [ r/(r+a_s) ]^2$.
We integrate the satellite orbit with the semi-analytical code described in \citet{2020MNRAS.494.3053B}, in which we incorporated the evolution of the tidal radius as detailed in Section \ref{sec:Tid_rad}. We truncate the satellite mass profile integrating $m_s(r)$ up to $R_t$.  The semi-analytical framework features a comprehensive treatment of the dynamical friction specifically tailored to account for flattened and rotating systems \citep{2020MNRAS.494.3053B, 2021MNRAS.502.3554B}. It is also equipped with a prescription for the interactions of massive perturbers with galactic substructures such as bars \citep{2022MNRAS.512.3365B}.

\subsection{Host galaxy}\label{sec:host}
In the present work, we explore two different models for the host galaxy: a single-component and a double-component host galaxy.
In the first scenario, the primary galaxy is characterised by an isolated exponential disc, defined by the density profile:
\begin{equation}
    \rho_d (R,z) = \frac{M_d}{4 \pi R^2_d z_d} \mathrm{e}^{-\frac{R}{R_d}} \, \mathrm{sech}^2 \bigg (\frac{z}{z_d} \bigg).
\label{eq:disc_rho}
\end{equation}
Here $M_d$ is the total mass of the disc, $R_d$ and $z_d$ are the scale radius and height of the disc, respectively.
An analytical approximate expression for the potential of such a model exists only within the galactic plane.
Consequently, accelerations caused by the disc potential outside the galactic plane are determined through numerical interpolation of tabulated values, which are computed over an adaptive grid, see \cite{2020MNRAS.494.3053B, 2021MNRAS.502.3554B} for details.
Single-component host galaxy models were employed to test simple systems, in which we neglect dynamical friction to focus on the tidal effects regulating the evolution of the satellite mass. 

In the case of a composite host galaxy, the disc is embedded within a spherical dark matter (DM) halo. The potential of this halo follows the Hernquist profile \citep{1990ApJ...356..359H}, characterised by a total mass $M_h$ and a scale radius $a_h$:
\begin{equation}
    \Phi_h(r) = - \frac{G M_h}{r+a_h}.
    \label{eq:pot_dm_halo}
\end{equation}
This choice is motivated by the fact that the Hernquist profile is numerically convenient and indistinguishable in the inner region from a Navarro Frank and White (NFW) \citep{1997ApJ...490..493N} profile. For this reason, it has been extensively used in literature to model DM halos \citep[see e.g.][]{2014MNRAS.444...62Y}.

\subsection{N-body simulations}
Our investigation was complemented by a comparative analysis, where we accompanied the proposed semi-analytical prescription regulating the tidal-stripping-driven mass evolution of satellite galaxies with N-body simulations. This approach enables us to evaluate the ability of our model to accurately encompass all the relevant physical processes involved and identify potential missing effects.
N-body simulations were performed employing the publicly available code \textsc{GADGET-4} \citep{2021MNRAS.506.2871S}

In all the tested systems, the satellite galaxy is modelled with $10^5$ stellar particles. The particle positions are initialised to follow the mass distribution in Eq.~\eqref{eq:hern_dens}, while the velocities are generated at equilibrium in the potential generated by the stellar distribution. 
 The initial satellite mass is fixed to be equal across all models, with $M_s = 10^8 M_{\odot}$, ensuring a sufficiently small satellite-to-host mass ratio to avoid significant perturbations on the host's potential, as we consider the latter fixed.
 We considered three different values for the satellite scale radius , i.e. $a_s = 0.1, \, 0.5, \, 1$ kpc, thus testing different mass concentrations.

\begin{table*}[ht]      

\caption{Parameters of the host galaxy for both single and the double component scenarios.}
\centering
\begin{tabular}{l c c c c c c c}
\hline\hline  

Component  &  & Profile & M  & scale radius & scale height & $N_{part}$ & $\epsilon$ \\
    \hline
    \noalign{\smallskip}
    \textbf{Single-component} & & & & \\
    
    Disc - analytical & & $\mathrm{exponential}$ &$ 4.4 \times 10^{10}\,  \mathrm{M_{\odot}}$ & $4.25$\,  $\mathrm{kpc}$ & $0.85 \, \mathrm{kpc}$ & - & - \\
    \noalign{\smallskip}
    \noalign{\smallskip}
    \noalign{\smallskip}
    \textbf{Double-component} & & & & \\
    Disc  & & $\mathrm{exponential}$ & $4.4 \times 10^{10}\,  \mathrm{M_{\odot}}$ & $4.25\,  \mathrm{kpc}$ & $0.85 \, \mathrm{kpc}$  & $10^7$ & $5 \, \mathrm{pc}$\\
    Halo - analytical &  & $\mathrm{Hernquist}$ & $1.1 \times 10^{12}\,  \mathrm{M_{\odot}}$ & $37\,  \mathrm{kpc}$ & - & - & -\\
\hline
\end{tabular}\label{tab:host_params}
\end{table*}
 
The satellite is then embedded within the primary galaxy at a distance of $R_i  = 10$ kpc from its centre and with a specific initial velocity, which is added to the stars as a bulk velocity. 
We explore the orbital parameter space by changing both the initial velocity of the satellite CoM ($v_i / v_c = 0.75, \, 0.50, \, 0.25 $, where $v_c$ is the circular velocity at $R_i$), and different initial inclinations of the satellite orbit with respect to the galactic plane ($\theta = 0^{\circ}, 30^{\circ}, 60^{\circ}, 90^{\circ}$). We set the softening parameter $\epsilon = 1$ pc for the satellite particles, while we fix $\epsilon = 5$ pc for the stellar particle of the disc component in multi-component galaxy models.
To isolate the impact of tidal forces on the evolution of the satellite mass from other possible influencing processes, we first performed a set of simulations excluding the effect of dynamical friction. To achieve this, the host galaxy is included in N-body simulations as a stationary semi-analytical potential, instead of being modelled using collisionless particles. To do so, we add to the acceleration of satellite particles the acceleration induced by the presence of the host potential. As mentioned in the previous section, all the models in which we omit dynamical friction host a primary galaxy modelled with a single exponential disc.

The method we implemented in \textsc{GADGET-4} to compute the accelerations generated by the exponential-disc potential is analogous to the one we use in the semi-analytical code and described in sec. \ref{sec:host}.
This set up prevents gravitational interactions between satellite and field stars, thereby avoiding dynamical friction to take place. 

We then consider more complex systems composed by a satellite orbiting in a double-component host galaxy, also including effects from dynamical friction.
In these systems, the primary galaxy consists of an analytical dark matter halo, whose potential is given by Eq.~\ref{eq:pot_dm_halo}, and an exponential-disc, modelled with $10^7$ stellar particles, whose mass density is given by Eq. ~\eqref{eq:disc_rho}. 
The initial conditions for the disc were performed using the public code \textsc{GalIC} \citep{2014MNRAS.444...62Y} which is based on an iterative approach to build N-body galaxy models at equilibrium.
Similarly to the case of the analytic disc, the dark matter halo contributes solely through the acceleration its potential imprints on the stellar particles - that we compute and add to the satellite particles in the simulation -, thus giving null contribution to the dynamical friction. 

The host galaxy parameters are summarised in Table~\ref{tab:host_params}.

\subsection{Satellite CoM and bound particles}
\begin{figure*}
   \centering
   \includegraphics[width=\hsize]{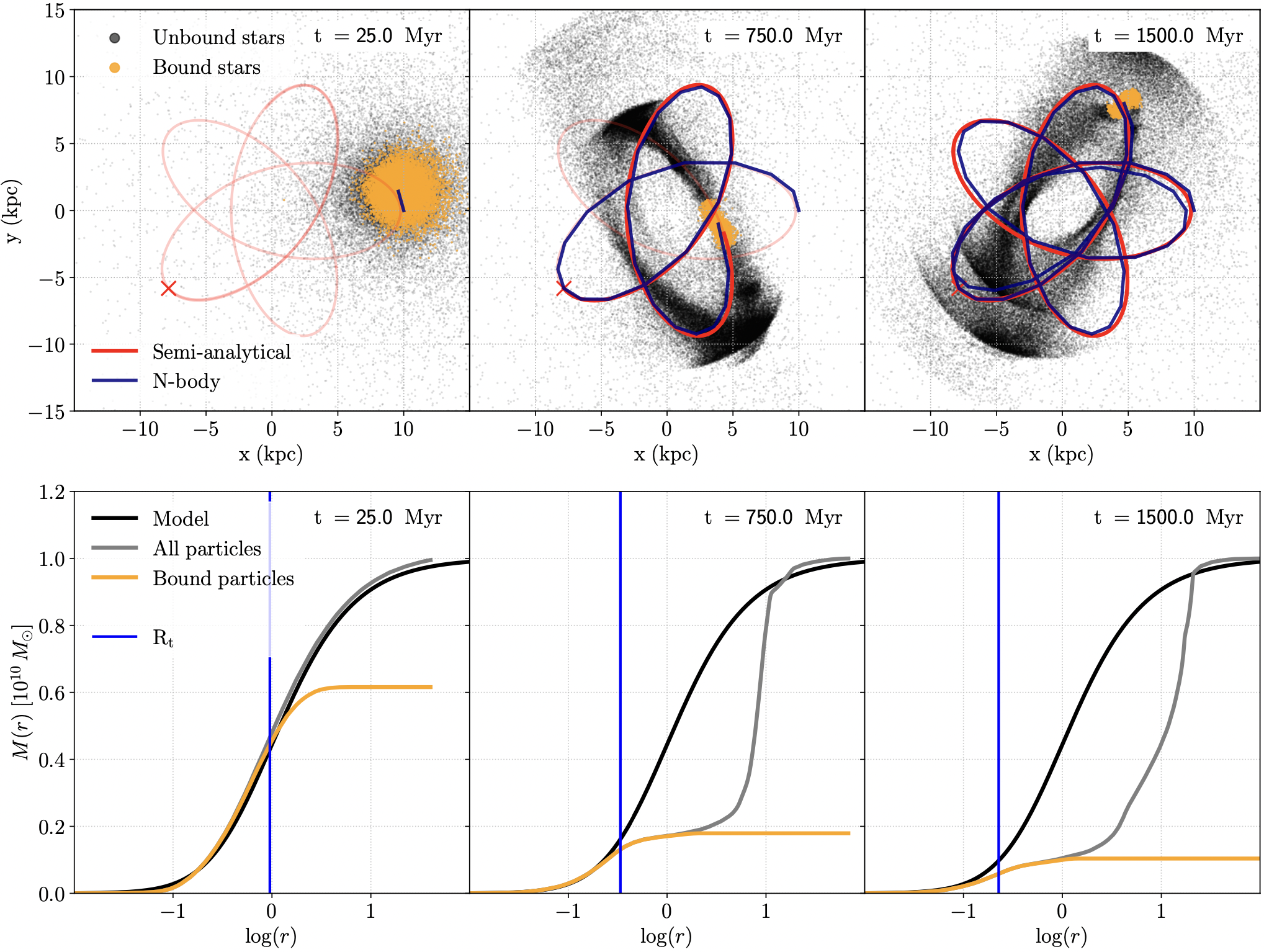}
   \caption{\textit{Upper panels}: satellite particles of an example run featuring a satellite with $a_s = 0.5$ kpc, orbiting in the galactic plane of an exponential disc host, with initial velocity $v_i = 0.5 \, v_c$. From left to right, the three panels correspond to the first, middle and final snapshot of the simulation. The origin coincides with the centre of the host galaxy potential. The colours indicate which particles are bound to the satellite (orange) or unbound (grey). The shaded red line shows the trajectory predicted by the semi-analytical model, while the solid red and blue lines track the satellite CoM, in the semi-analytical model and in the N-boy simulation, respectively. Finally, the red cross indicates the initial point for the semi-analytical orbital integration, corresponding to the first apocentre. \textit{Lower panels}: satellite cumulative mass profiles at the same snapshots and for the same system as in the upper panels. The black solid curve displays the theoretical cumulative mass profile from the Hernquist model. The other two profiles are constructed using the bound particles only, in orange, and all the particles that were part of the satellite at the initial time, in grey. The vertical blue line shows the value of the tidal radius computed with our semi-analytical prescription.
   }
    \label{Fig:orbit_mass_snap}
    \end{figure*}

The upper panels in Fig.~\ref{Fig:orbit_mass_snap} show satellite particles in one of the tested models (specifically the system composed of a satellite with $a_s = 0.5$ kpc, orbiting in the galactic plane of an exponential disc host, with initial velocity $v_i = 0.5 \, v_c$) at the first, middle and final snapshot of the simulation. The plots' origin coincides with the centre of the host galaxy potential. Orange particles are bound to the satellite, while grey particles indicate those that have been stripped. The shaded thin red line shows the trajectory predicted by the semi-analytical model, while the thick solid red and blue lines track the satellite CoM, in the semi-analytical model and in the N-boy simulation, respectively.
In each snapshot of the simulation the bound particles are identified through an iterative approach.

We start by identifying the position and velocity of the satellite CoM. We initialise the satellite CoM location as the point corresponding to the highest density. For each of the satellite particles we compute the binding energy as:
\begin{equation}
    E_*  = \frac{1}{2}|\mathbf{v_*}-\mathbf{v_{\rm CoM}}|^2 - \Phi_{\rm Trunc \, Hern}(r_*).
\end{equation}
Here $v_*$ is the velocity of the star, $v_{\rm CoM}$ is the velocity of the satellite CoM, and $\Phi_{\rm Trunc \, Hern}(r_*)$ is the potential generated by an Hernquist model, truncated at a certain radius $r_{\rm max}$, which is given by:
\begin{equation}
    \Phi_{\rm Trunc \, Hern}(r_*) = 
        \begin{cases}
            G M_s \big ( \frac{1}{r_{\rm max}+a_s}- \frac{1}{r_{\rm max}} - \frac{1}{r_*+a_s} \big )  &\quad \mathrm{if} \; r_* < r_{\rm max} \\
            - \frac{G M_s}{r_*} &\quad \rm{if} \; r_* \geq \ r_{\rm max} \\
        \end{cases}\; ,
\end{equation}
where $r_*$ is the distance of the selected star from the satellite centre.
To determine the truncation radius $r_{\rm max}$ at each snapshot, we initially set $r_{\rm max} = 10 a_s$, and subsequently, we consider enlarging spherical shells centred at the satellite CoM with a fixed width of $\delta_r = 0.25 a_s$. The value of $r_{\rm max}$ is then chosen to correspond to the median radius of the smallest shell containing a number of unbound stars exceeding twice the number of the bound ones  (i.e. such that $N_{\rm unbound} \geq 2  N_{\rm bound}$). 
We update the CoM location and velocity with the values computed using the stars with $E_* <0$. The procedure is iteratively repeated until the CoM position converges to a constant point, with a relative error on the position of the CoM lower than $10^{-3}$.

The lower panels in Fig. \ref{Fig:orbit_mass_snap} show the satellite cumulative mass profile at the same snapshots and for the same system as in the upper panels. The black solid curve displays the theoretical cumulative mass profile from the Hernquist model. The other two profiles are constructed using the bound particles only, in orange, and all the particles that were part of the satellite at the initial time, in grey. The vertical blue line shows the value of the tidal radius computed with our semi-analytical prescription at the same time of the simulation. Thus, the satellite mass resulting from the simulation, given by the value at which the orange curve saturates, can be compared to the value predicted by our semi-analytical model, i.e. the value at which the theoretical profile is truncated by the tidal radius.

\subsection{Mass evolution and choice of the optimal $\alpha$ parameter}
\label{sec:alpha_best}
\begin{figure*}
   %\centering
   \includegraphics[width=\hsize]{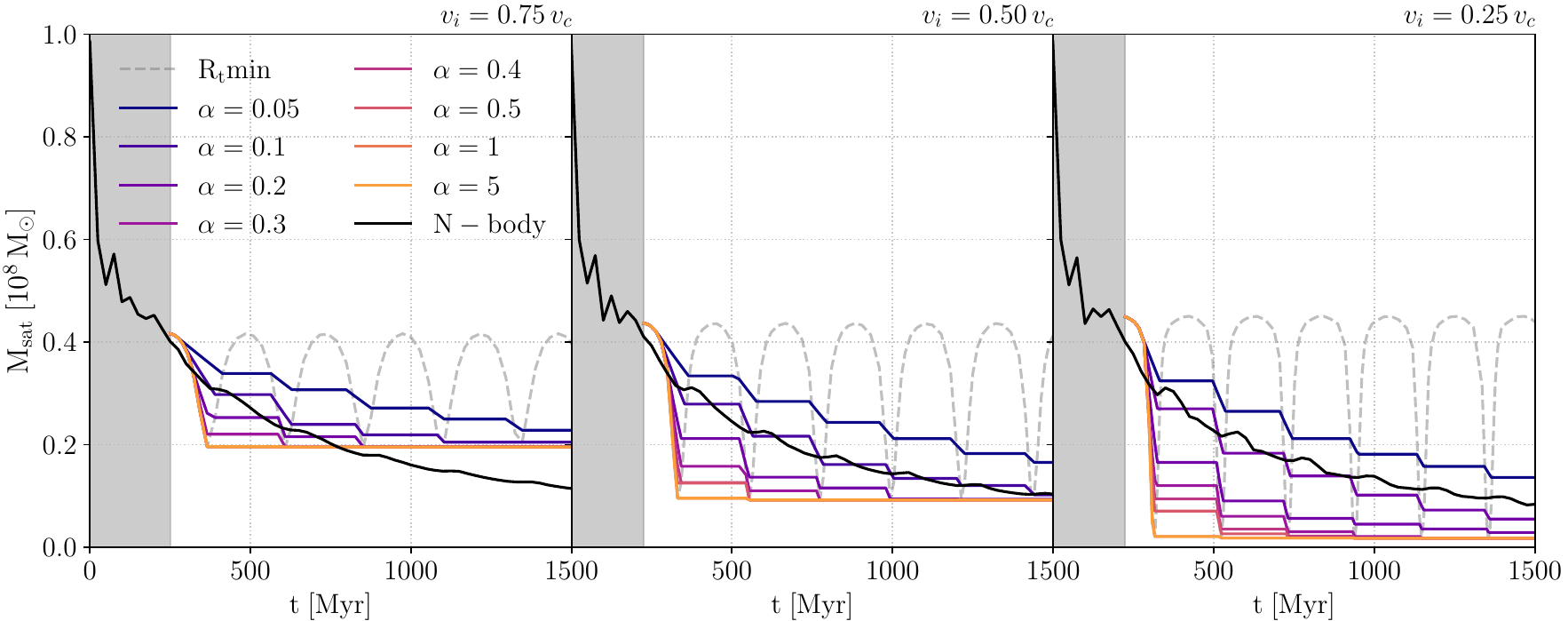}
   \caption{Evolution of the satellite mass as a function of time for three cases with $a_s =0.5$ kpc, orbiting within the host galactic plane and featuring different initial velocities ($v_i/v_c = 0.75 \, , 0.5 \, , 0.25$, from left to right). The black line shows the evolution of the mass according to the N-body simulations. The coloured solid lines correspond to the mass evolution predicted by the semi-analytical model with different values for $\alpha$, spanning the range $[0.05-5]$. The grey dashed line represents the mass predicted using the minimum tidal radius  evaluated along the 1000 different directions - if let free to increase -, whereas the grey region represents the time from the beginning of the N-body simulation to the first apocentre, which is the starting point for the semi-analytical models.}
              \label{Fig:mass_evol_IP}%
    \end{figure*}

We compare outcomes of N-body simulations with the results of our semi-analytical prescription, testing different values of the $\alpha$ parameter, which controls the mass-stripping rate. A higher $\alpha$ corresponds to a faster mass removal.
The panels in Fig. \ref{Fig:mass_evol_IP} illustrate the temporal evolution of the satellite mass of a satellite with $a_s =0.5$ kpc orbiting within the host galactic plane, for three different initial velocities: $v_i/v_c = 0.75 \, , 0.5 \, , 0.25$. The black line shows the evolution of the mass resulting from N-body simulations.  
The coloured solid lines display the mass evolution predicted by the semi-analytical model for different values of $\alpha$, spanning from $0.05$ up to $5$. The minimum tidal radius computed at each time is indicated with a grey dashed line, which indicates the value of the satellite mass one would predict if the stripping were considered instantaneous and reversible.
It is important to notice that the initial configuration of the simulated systems is not at the equilibrium. This is because the satellite is generated in isolation and then artificially placed within the primary galaxy potential, instead of following the merger from its initial phases. Therefore, we use the position and velocity of the satellite CoM in the N-body simulation at the apocentre after the first orbit as the initial condition for the semi-analytical model calculations. In Fig. \ref{Fig:mass_evol_IP}, the first orbit is indicated by the grey shaded region.
Finally, using a least square method on the mass evolution, we determine the optimal value of $\alpha$ corresponding to the semi-analytical model that most accurately reproduces the N-body simulations.

Importantly, to make sure that our results are not affected by artificial numerical stripping, we compared the outcomes of our simulations with the criteria proposed in  \cite{van_den_Bosch_2018}\footnote{The criteria in \cite{van_den_Bosch_2018} are computed assuming a Navarro-Frenk-White \citep{1997ApJ...490..493N} profile for the satellite galaxy. We applied these criteria to our satellites, even though our analysis employs a Hernquist model. Extending the computation to determine the precise threshold for a Hernquist profile is beyond the scope of this paper.}. The number of particles ($N = 10^5$) and the small softening length ($\epsilon = 1$ pc) used to model our satellite galaxies place our results well above (about two orders of magnitude) the threshold ensuring that the system does not suffer from both discreteness noise and inadequate force resolution in all the tested cases and over the entire simulation time.

In the next section, we will discuss the results of our model, focusing in particular on the model ability to reproduce the evolution of the satellite mass.

\section{Results}
\label{sec:results}

\subsection{Models without dynamical friction}
\label{Models_without_DF}

To test the efficiency of our semi-analytical model in predicting the evolution of a satellite galaxy within a non-spherical host, we started our investigation by considering the limiting case of a satellite moving in the analytical potential of a single-component disk-like host galaxy. Although far from being realistic, this configuration allows us to isolate the effects determined by tidal forces exerted only by the disc, excluding the influence of other factors that can affect its orbital evolution, such as the presence of a spherical component in the host galaxy and the effect of dynamical friction.

\begin{figure*}
\centering
\includegraphics[width=\hsize]{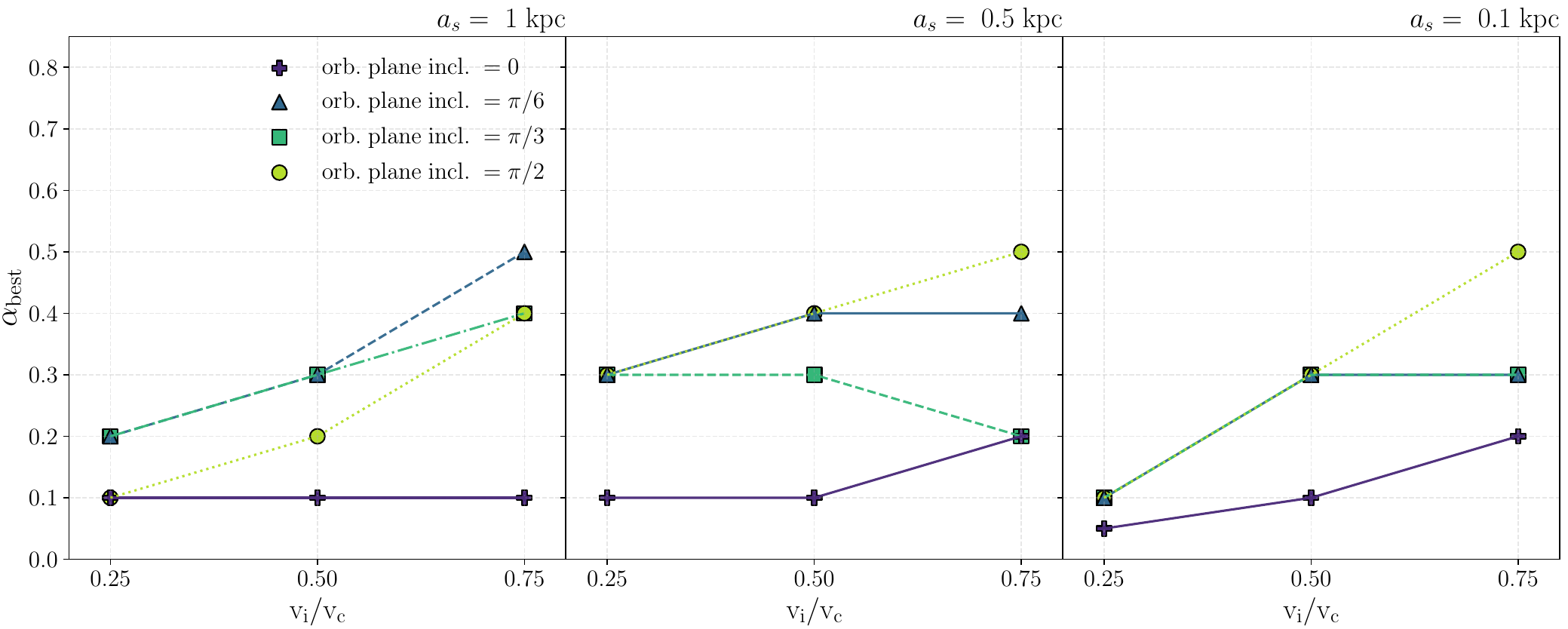}
\caption{Optimal values of the $\alpha$ parameter for each model without dynamical friction as a function of the initial orbital velocity (or initial eccentricity). Each panel considers a system with fixed satellite scale radius $a_s$, decreasing from left to right, with varying linestyles and color codes indicating different orbital inclinations.}   \label{Fig:alpha_best_NoDF}%
\end{figure*}

Fig.~\ref{Fig:alpha_best_NoDF} displays the optimal values of the $\alpha$ parameter for each model, evaluated as detailed in sec.~\ref{sec:alpha_best}. More in detail, the three panels show how the $\alpha_{\rm best}$ parameter changes with the initial orbital velocity (or initial eccentricity) in models sharing the same satellite scale radius $a_s$, each panel referring to a different value of $a_s$, and the same orbital inclination, reported with different line styles and colours. In general, most systems exhibit a slight increase in the $\alpha$ parameter as the initial velocity approaches the circular velocity, while no evident trends in the values of $\alpha$ can be outlined when varying the scale radius and the orbital inclination. As expected, a lower $\alpha$ is associated to systems with initial higher eccentricity (or lower initial velocity). This is attributed to the abrupt decrease in the tidal radius at pericentre passages, as predicted by Eq.~\eqref{Eq:new_def_TR}, leading to a significant and instantaneous mass loss. However, the actual timescale to strip material from the satellite, as predicted by N-body simulations, is longer than the fast pericentre passages. For this reason, in the vicinity of the pericentre, the tidal radius decrease is delayed using Eq.~\ref{eq:TR_2}, with $\alpha$ regulating the rapidity of the mass removal. Since this effect is much more relevant along eccentric orbits, the $\alpha$ parameter needs to be small enough to slow down the satellite mass loss, which otherwise would be extreme, and is expected to be smaller compared to systems with low eccentric orbits.
If not explicitly specified, all the results presented from this point forward refer to the specific semi-analytical model characterised by the optimal value of $\alpha$ for each system considered.
 
   \begin{figure*}
   \centering
   \includegraphics[width=\hsize]{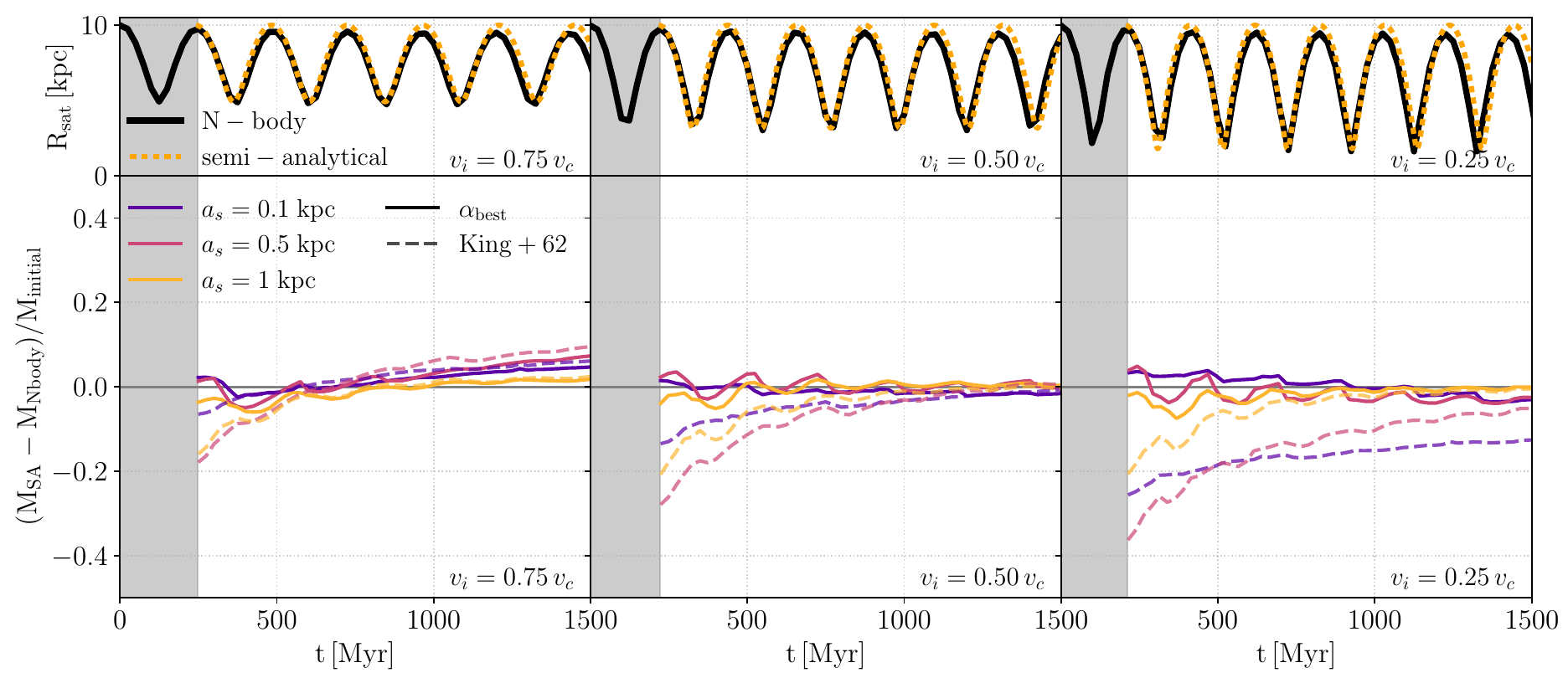}
   \caption{Results obtained for systems hosing satellites orbiting within the galactic plane without dynamical friction. The three panels refers to different initial velocities for the satellite CoM, $v_i = 0.75$, $v_i = 0.5$ and $v_i = 0.25$ form left to right.  \textit{Upper panels}: separation of the satellite CoM from the primary galaxy center as a function of time. The thick orange dotted line refers to the semi-analytical model, while the thick black solid line shows the result of the N-body simulations. \textit{Bottom panels}: time-evolution of the difference between the satellite mass predicted by the semi-analytical model and the mass resulting from N-body simulations, normalized to the initial satellite mass. The line colors indicate different satellite scale radii. The solid lines refer to our new semi-analytical prescription for the evolution of the satellite mass, whereas the dashed lines represent the results we obtain using King's formula for the tidal radius. In both panels, the gray area indicates the time interval leading to the first apocenter. }
              \label{Fig:IP_ba_k}%
    \end{figure*}

In Fig. \ref{Fig:IP_ba_k}, we present the results of the comparison between our semi-analytical prescription and N-body simulations for models with the satellite moving within the galactic plane. The upper panels depict the evolution of the separation of the satellite CoM from the primary galaxy centre. The semi-analytical model's predictions are shown in orange, while the N-body simulation results are represented by a black solid line. The bottom panels show the time evolution of the difference between the satellite mass (normalised to the initial satellite mass) predicted by the semi-analytical model and the mass resulting from N-body simulations. The three panels correspond to different initial velocities of the satellite, with line colours indicating the satellite scale radius.
Our semi-analytical prescription well reproduces both the orbital and the mass evolution of the satellite. 

As an additional test, we compare our semi-analytical prescription for the tidal radius and mass evolution (solid lines) with results obtained using King's formula (dashed lines), see Eq.~\eqref{Eq:TR_king}.
We observe an overall better agreement with N-body simulations using our new semi-analytical prescription compared to the King prescription. This result is due to multiple factors.
First, King's formula, when applied without any delay for mass removal, implies instantaneous mass stripping. This leads to a general underestimation of the satellite mass, especially in the initial phases of the evolution. Moreover, one of the main assumptions in King's prescription is that the tidal radius should be much lower than the separation between the centres of the two galaxies, thereby excluding close encounters. This assumption is generally valid along quasi-circular orbits, but it breaks when considering highly eccentric orbits where the pericentre can be at a close distance from the host centre. The combined effect of the instantaneous mass stripping, which can be severe in eccentric orbits during the close pericentre passages, and the assumption of distant interactions, imply an increasing inability of King's prescription at reproducing the results of N-body simulations (see bottom right panel in Fig.~\ref{Fig:IP_ba_k}). 

It is important to note that a comparison with  King's prescription is meaningful only for systems in which the satellite is orbiting within the galactic plane, as far from the galactic plane King's definition of the tidal radius becomes ill-defined. In the co-planar case, indeed, the gradient of the host potential at the position of each satellite's star points approximately toward the host centre, making the comparison between our and King's prescriptions meaningful. Nonetheless,  we stress that, even in this case, the acceleration of stars that during their orbits around the satellite centre lie above or below the plane of the host disc are not radial, and are, therefore, implicitly approximated in the treatment by \cite{1962AJ.....67..471K}. 
 
Finally, we investigated systems where the satellite orbits outside the galactic plane, exploring various inclination angles. Since the qualitative trends observed in these cases are similar to the ones discussed for co-planar orbits, we show the evolution of the error in estimating the satellite mass for these systems in Fig. \ref{Fig:Mass_err_OP_APO}.

Our semi-analytical prescription effectively reproduces the evolution of the satellite mass along the orbit, particularly in systems with eccentric orbits, across all orbital inclinations. However, in systems hosting satellites with low-eccentricity orbits, our semi-analytical model tends to overestimate the satellite mass, as observed in the right panels of Fig.s~\ref{Fig:IP_ba_k} and \ref{Fig:Mass_err_OP_APO}. We will delve into this behaviour extensively in Section \ref{sec:low-ecc, tan vel}.

\subsection{Models with dynamical friction}   

After assessing the capability of our model to replicate the effects of tidal stripping in a fixed analytical potential, we extend our analysis to include models where dynamical friction is considered. In this context, our study involves satellite galaxies orbiting within a multi-component host galaxy. As detailed in Table~\ref{tab:host_params}, the host galaxy in these models comprises a spherically symmetric dark matter halo, incorporated as an analytical potential in N-body simulations, and an exponential disk containing $10^7$ stellar particles. Consequently, the dynamical friction experienced by the satellite stars is solely attributed to the disk component of the host galaxy. 
In contrast to the models examined thus far, the introduction of dynamical friction, as described in detail in the introduction, significantly influences the satellite's orbital evolution, which, in turn, plays a crucial role in shaping the tidal radius and consequently determining the extent of mass removal.

   \begin{figure*}
   %\centering
   \includegraphics[width=\hsize]{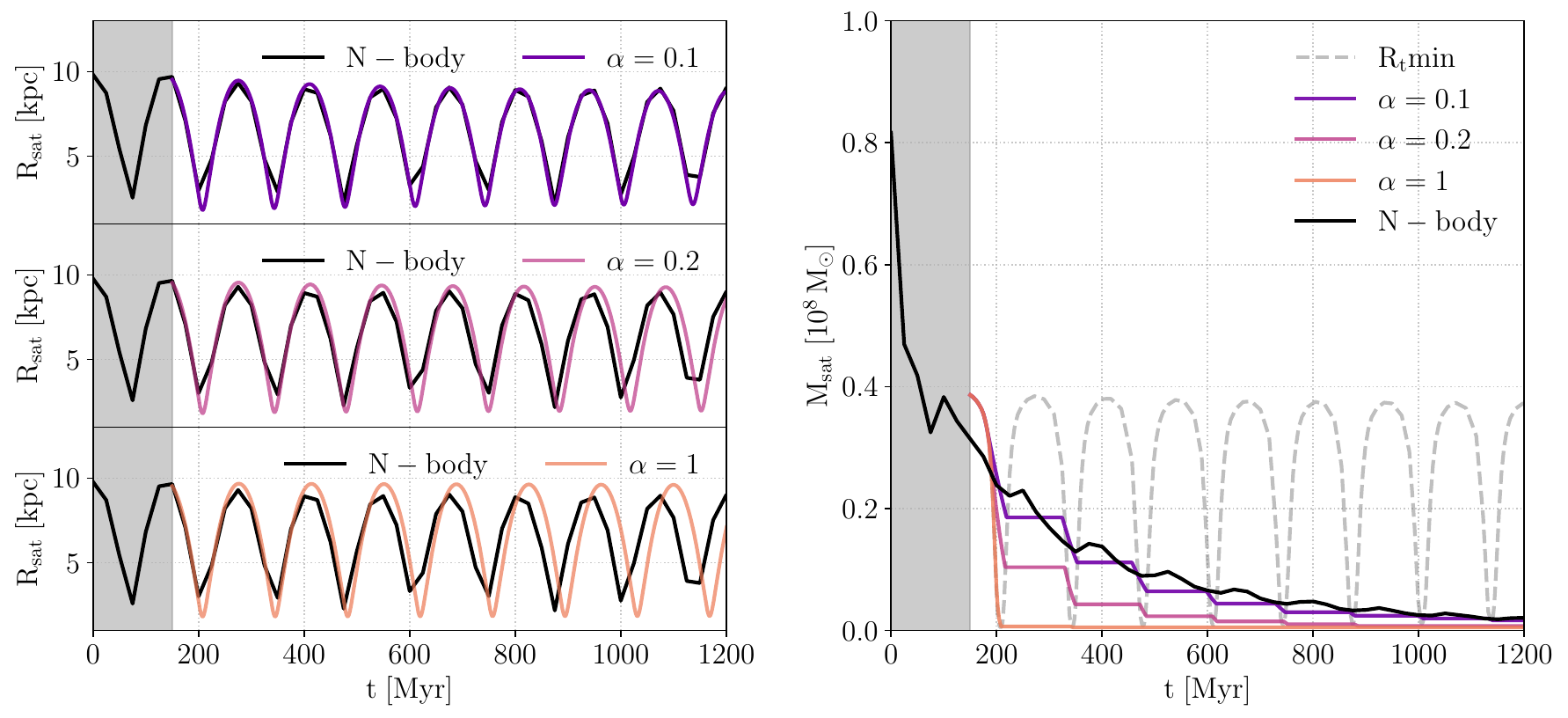}
   \caption{\textit{Left panels}: time-evolution of the satellite's distance from the centre of the host in the N-body simulation (black lines) compared to our semi-analytical model's predictions (coloured solid lines). From top to bottom, each panel refers to different values of $\alpha$:  0.1,  0.2, and  1, respectively. \textit{Right panel}: mass evolution of the satellite in both N-body simulations and semi-analytical models, maintaining the same colour code as in the left panels. The grey dashed line represents the mass predicted using the minimum tidal radius  evaluated along the 1000 different directions - if let free to increase.}
              \label{Fig:M_orb_DF}%
    \end{figure*}

The combined effect of dynamical friction and mass loss is illustrated in Figure \ref{Fig:M_orb_DF}, where we report the result for one of the systems we tested (i.e. a satellite orbiting within the galactic plane with initial velocity of $v_i = 0.25 v_c$ and $a_s=0.5$ kpc \footnote{Due to the computational cost of simulations involving a high number of particles, and since the results of simulations without DF are almost independent of the satellite scale radius, we chose to consider a single value for $a_s$. We picked $a_s = 0.5$ kpc, i.e. the middle value among those that we tested in the previous sections.}). The left panels compare the satellite's distance evolution from the centre of the host in the N-body simulation (depicted by the black line) with our semi-analytical model's predictions for three distinct $\alpha$ values (each represented by a coloured solid line in a separate panel). Correspondingly, the right panel shows the satellite's mass evolution in both N-body simulations and semi-analytical models, maintaining the same colour code as in the left panels.

In the right panel of fig. \ref{Fig:M_orb_DF}, similarly to fig. \ref{Fig:mass_evol_IP} and fig. \ref{Fig:M_evol_VTAN}, it is possible to notice small increases in the satellite mass, occurring just after pericentre passages. Those increases are due to satellite particles that are stripped during the pericentre passage but, thanks to their orbital motion, are re-accreted soon after the closest approach to the host centre, rebinding to the satellite. The amount of matter re-accreted is very small compared to the amount of matter that one would predict to rebind to the satellite after each pericentre passage in the case of a freely evolving $R_t$ (grey dashed line). For this reason, and also the fact that the amplitude of this bump in the satellite mass gets damped with the subsequent pericentre passages, we neglect this effect and consider the $R_t$ to be a decreasing function of time. 

Among the models investigated, the one corresponding to $\alpha = 0.1$ exhibits the best agreement with both the satellite's mass and orbital evolution. Conversely, models associated with higher values of $\alpha$, corresponding to faster mass loss, demonstrate an increasing deviation from simulations results. This discrepancy arises from the rapid reduction in the satellite mass, which leads to a weakening of the dynamical friction drag,
consequently slowing down the satellite's decay towards the host centre.

The best values of the $\alpha$ parameter for all the investigated systems are summarised in Table~\ref{tab:best alpha}.
As highlighted in the previous section, models devoid of dynamical friction exhibit a consistent agreement between our semi-analytical model and N-body simulations, independently of the scale radius and orbital inclination, with a mild dependence on the initial orbital eccentricity only. Given this result, and the fact that 
simulations involving a host disk composed of $10^7$ particles represent a quite high computational burden compared to simulations with entirely analytical hosts, we opt to focus our investigation on systems featuring a satellite with a fixed scale radius, $a_s = 0.5$ kpc, orbiting within the galactic plane. The primary parameter under consideration is therefore the variation in the satellite's initial velocity.

\begin{figure*}
   %\centering
   \includegraphics[width=\hsize]{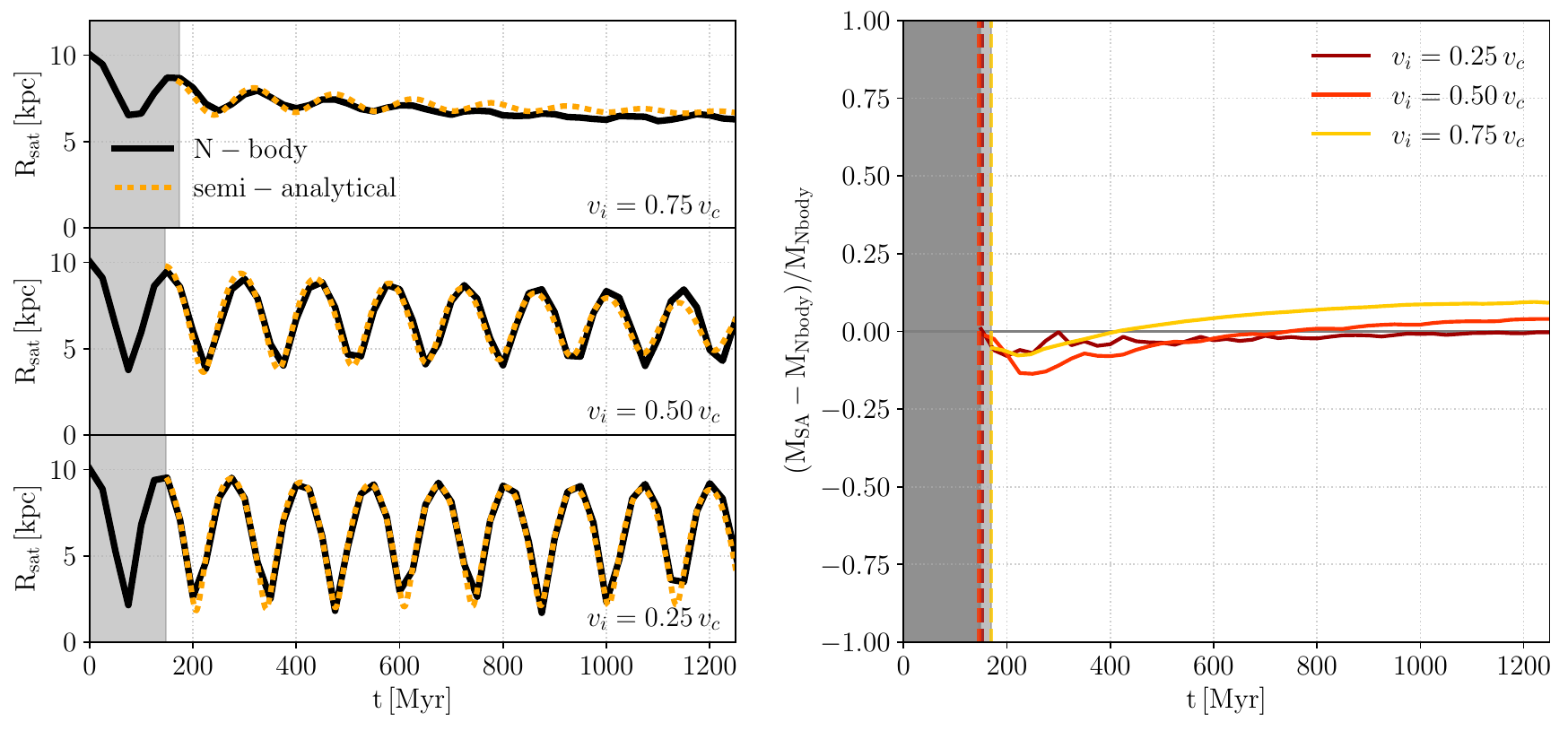}
   \caption{\textit{Left panels}: comparison between the evolution of the satellite's CoM in both N-body simulations and semi-analytical models, each using the best value for $\alpha$. The different panels correspond to the three different initial satellite's velocities, $v_i = 0.75 v_c$,  $v_i = 0.50 v_c$ and $v_i = 0.25 v_c$ from top to bottom. The thick orange dotted line refers to the semi-analytical model, while the thick black solid line shows the result of the N-body simulations. \textit{Right panel}: relative error in the evaluation of the satellite mass for the same values of the initial velocities as a function of time. Different line colours indicate different initial satellite velocities. The dashed vertical lines represent the initial time of the semi-analytical models, which corresponds to the first apocentre, and are coloured using the same colour code employed for the solid lines.}
              \label{Fig:Mass_err_DF}%
    \end{figure*}

The results are shown in Fig.~\ref{Fig:Mass_err_DF}. The left panels compare the evolution of the satellite's CoM in both simulations and in semi-analytical models, each using the best value for $\alpha$. From top to bottom, the different panels correspond to the three different initial satellite's velocities, $v_i = 0.75 v_c$,  $v_i = 0.50 v_c$ and $v_i = 0.25 v_c$. The right panel depicts the error in the evaluation of the satellite mass for the same values of the initial velocities. The dashed vertical lines represent the initial time of the semi-analytical models, which correspond to the first apocentre, and are coloured using the same colour code as the solid lines.

As noted in the previous cases, a very good agreement is observed between the results obtained from N-body simulations and the predictions from our semi-analytical models regarding the orbital evolution of the satellite and the associated mass decrease. Notably, this accord is particularly pronounced for systems featuring satellites on higher eccentric orbits, as consistently demonstrated across all the investigated systems.

\begin{table}[htb]
\caption{Values of the $\alpha$ parameter for each model in this work.}
\renewcommand{\arraystretch}{1.3}
\begin{tabular}{c|c|c|c|c}

\multirow{2}{*}{$v_i/v_c$} & \multirow{2}{*}{$\theta$} & \multicolumn{3}{c}{$a_s$} \\
\cline{3-5}
& & 0.1 \rm kpc & 0.5 \rm kpc & 1 \rm kpc \\

\hline
\hline
\textbf{No Dynamical friction} & & & \\
\multirow{4}{*}{0.75} & 0 & 0.2 & 0.2  & 0.1   \\
                      & $\pi/6$ & 0.3 & 0.2  & 0.5   \\
                      & $\pi/3$ & 0.3 & 0.2  & 0.4   \\
                      & $\pi/2$ & 0.5 & 0.5  & 0.4   \\
                     
\hline
\multirow{4}{*}{0.50} & 0 & 0.1 & 0.1  & 0.1   \\
                      & $\pi/6$ & 0.3 & 0.4  & 0.3   \\
                      & $\pi/3$ & 0.3 & 0.3  & 0.3   \\
                      & $\pi/2$ & 0.3 & 0.4  & 0.3   \\
\hline                
\multirow{4}{*}{0.25} & 0 & 0.05 & 0.1  & 0.1    \\
                      & $\pi/6$ & 0.1 & 0.3  & 0.2   \\
                      & $\pi/3$ & 0.1 & 0.3  & 0.2   \\
                      & $\pi/2$ & 0.1 & 0.3  & 0.1   \\
\hline
\hline
\textbf{Dynamical friction} & & & \\
0.75& 0 & - & 0.1  & -    \\
0.50& 0 & - & 0.1  & -    \\
0.25& 0 & - & 0.1  & -    \\

\hline
\end{tabular}

\label{tab:best alpha}
\end{table}

\subsection{Testing low-eccentricity satellite orbits} 
\label{sec:low-ecc, tan vel}

In this section, we investigate in detail the processes contributing to the systematic overestimation of satellite mass in our semi-analytical model when compared to N-body simulations in systems harbouring satellites on low-eccentricity orbits.
Two primary processes may account for this discrepancy. The first involves tidal heating resulting from rapid changes in the host potential experienced by the satellite, as described at the beginning of the methods. Another possible factor is the satellite's evaporation induced by mass truncation. During pericentre passages, where the majority of stripping occurs, a substantial portion of the satellite mass is expelled from the system, leading to truncation in the satellite mass distribution. As a result, the satellite is no longer in equilibrium. As it evolves towards a new equilibrium, its mass distribution expands, causing stars with higher velocities to migrate to larger radii. As a consequence, the satellite's profile changes becoming less concentrated, thereby facilitating the particles in the outer layers to become unbound. This results in a continuous mass loss, even if the tidal radius undergoes minimal change, particularly along quasi-circular orbits.

In order to discern the predominant process influencing the excess mass loss in the satellite, we conducted additional N-body simulations without dynamical friction. This was done to exclude potential additional effects that could contribute to the removal of mass from the satellite. The simulations were executed considering only systems characterised by the lowest initial orbital eccentricity, specifically with $v_i = 0.75 v_c$ , as these are the most affected by the process under investigation. The satellite under consideration featured a Hernquist mass distribution with $a_s = 0.5$. Instead of randomly oriented velocities, we initialised stars in the satellite on perfectly circular orbits, ensuring that no net rotation was imparted to the satellite as a whole.

To deal with the tendency of the velocities of the satellite stars to re-isotropise, a reorientation of the particles' velocities along the tangential direction was performed at every apocentre. Importantly, this reorientation did not alter the magnitude of the velocity vector, thus keeping the energies of the stars unchanged. This approach prevents stars on radial orbits from rapidly migrating towards larger radii, thereby restraining the overall evaporation of the satellite. This approach enables the discrimination between the  processes driving the excess in satellite mass loss. If the dominant factor is satellite evaporation, this methodology allows to reproduce the satellite mass evolution. Alternatively, if tidal heating is the primary driver, injecting energy into the satellite and causing the stars to acquire sufficient energy to escape the system, our simulation will still exhibit an excess in mass loss.

\begin{figure*}
   \centering
   \includegraphics[width=\hsize]{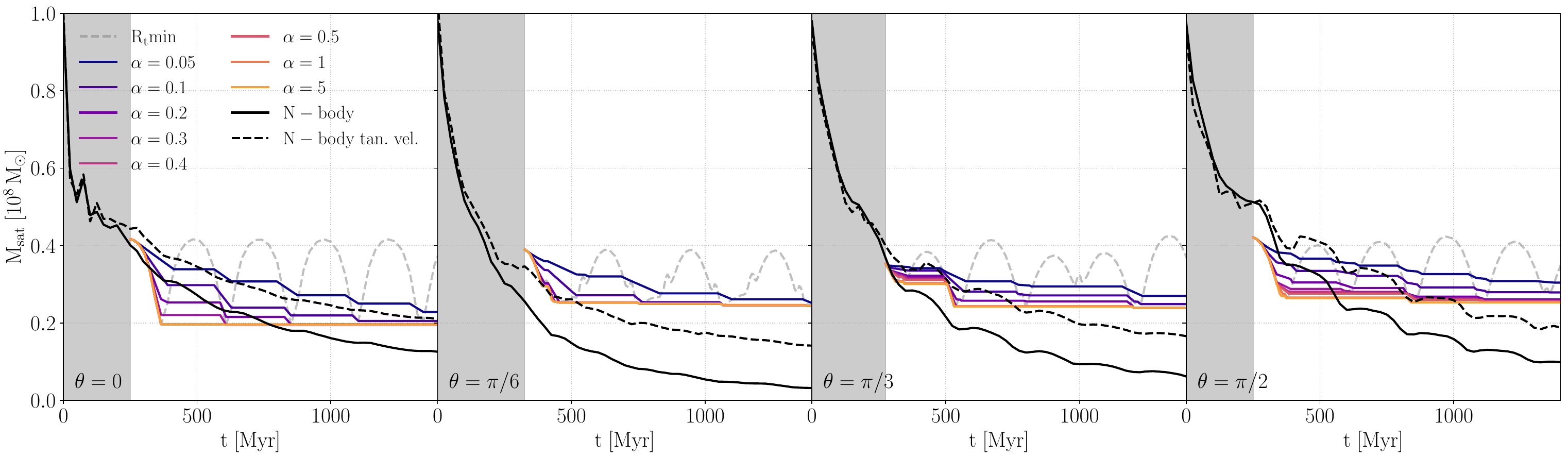}
   \caption{satellite mass as a function of time for systems with the reorientation of the satellite star velocities at the apocentres. The four panels refer to different orbital inclinations, from $\theta = 0$ (leftmost panel) to $\theta = \pi/2$ (rightmost panel). The black dashed line represents the satellite mass obtained through the new N-body simulations, compared with the outcome of the original N-body simulations showed as a black solid line. The coloured lines indicate the predictions of our semi-analytical model for various values of $\alpha$. The grey dashed line represents the mass predicted using the minimum tidal radius  evaluated along the 1000 different directions - if let free to increase. }
              \label{Fig:M_evol_VTAN}%
    \end{figure*}

The results are shown in Fig. \ref{Fig:M_evol_VTAN}. Each panel illustrates the satellite mass as a function of time for distinct orbital inclinations. The black dashed line represents the satellite mass obtained through the new N-body simulations, compared with the outcomes of the original N-body simulation presented in sec. \ref{Models_without_DF}, displayed as a black solid line. The coloured lines depict the predictions of our semi-analytical model for various values of $\alpha$.

In all systems, a substantial reduction in the mass loss rate is observed. Notably, the system harbouring a satellite orbiting within the galactic plane exhibits a satellite mass evolution now compatible with our semi-analytical model, particularly for $\alpha=0.05$. Conversely, in systems with orbits outside the galactic plane, although the reduction in satellite mass is more gradual compared to the original N-body runs, the stripped mass still exceeds that predicted by the semi-analytical models. This suggests that, at least within the galactic plane, the reorientation of star velocities is sufficient to reconcile the evolution with the semi-analytical model, indicating the dominance of satellite evaporation in shaping the mass evolution. Outside the galactic plane, however, tidal heating effects become significant, due to the stronger vertical gradient of the gravitational field in the proximity of the disk plane, and therefore it cannot be neglected.

\section{Discussion and conclusions}
\label{sec:conclusions}

In our analysis we evolved a satellite galaxy within a fixed (or quasi-fixed, for the simulations with live primaries) host potential. However, galaxies experience morphological evolution throughout cosmic time due to secular evolution. This evolutionary process may result from interactions between the galaxy and its environment, such as gas accretion or galaxy harassment, or it can be initiated by internal factors such as the presence of spiral arms or bars. Analysing cosmological simulations \cite{2023MNRAS.518.1427S,2024MNRAS.527.8841S}  showed that the growth of galaxies and their dark matter halos on sub-Gyr scales can significantly impact the evolution of merging satellite galaxies, especially affecting the satellite orbit during the pairing phase and, consequently, its infall time. Such a result is indeed backed-up by analytical arguments, such as those described in \cite{2020MNRAS.498.2219V}. Interestingly, and contrary to what is commonly expected, \cite{2023MNRAS.518.1427S,2024MNRAS.527.8841S} found that the satellite orbit is not always shrinking. Instead, some satellites exhibit an increase in the pericentre distance, often accompanied by a rise in the orbital specific angular momentum. This suggests that the growth of the host galaxy halo may promote the satellite migration to larger orbits, thus exerting a strong influence on its evolution. In light of these considerations, we have started applying our model to galaxies undergoing significant evolution, thus relaxing the constraint of a static primary galaxy potential dictating the motion of the satellite and allowing for both galaxies to evolve over time and pair together. The results of this analysis will be discussed in a forthcoming study.

In our study, we focused on the stellar and DM component of the merging galaxies, while we neglect the presence of gas in the merging galaxies. Our choice is motivated by our primary goal consisting in the characterisation of the tidal stripping of satellite galaxies in non-spherical hosts rather than a full inclusion of the different galactic components.  Nevertheless, concerning minor mergers, cosmological simulations show that these events can involve gas-rich satellites interacting with the gaseous component of their host \citep[see e.g.][]{2022NatAs...6..496M}. In such cases, the effect of ram pressure and of non-axisymmetric torques on the gas component represent  crucial mechanisms impacting the DF efficiency. On one side, by removing mass from the satellite galaxy \citep{1999MNRAS.308..947A,2006MNRAS.369.1021M,2022MNRAS.514.5276S,2023MNRAS.525.3849S}, ram-pressure slows down the satellite orbital evolution. On the other hand, \cite{Callegari_2009} showed that gas inflows triggered by non-axisymmetric structures\footnote{Similar inflow can be triggered by the ram pressure torques as well \citep[see, e.g.][]{2017MNRAS.465.2643C, 2018MNRAS.479.3952B}.} driven by the merger process stabilise the satellite nucleus against tidal disruption, leading to the successful completion of MBH pairing in unequal (1:10) galaxy mergers, while similar gas-free simulations resulted in the wandering of the smallest MBH at kpc scales. We plan to address these effect on the efficiency of DF and tidal stripping in gas-rich mergers in a future study.

Finally, we performed our study considering a single satellite-to-host mass ratio $<1:100$. Increasing the mass ratio would introduce significant distortions in the host potential, thus requiring dedicated studies and simulations which accounts for variations in the host potential and mass distribution.

In thiìs paper, we propose a new semi-analytical prescription for the tidal radius and the relative mass evolution of satellite galaxies in minor mergers. The novelty of the proposed approach primarily lies in the generalisation of the definition of the tidal radius to be suitable for any geometry and composition of the host galaxy, in contrast with traditional definitions \citep{1962AJ.....67..471K} which are provided for circular orbits, under the assumption of a spherical host. The prescription also accounts for a delay in the mass stripping and allows for eccentric orbits.

We validated our prescription against N-body simulations. In order to isolate the effects of tidal forces, we first consider systems not affected by dynamical friction, by considering a spherically symmetric satellite orbiting within the analytical potential of an exponential-disk host. We explored the parameter space by considering different initial orbital velocities, orbital inclinations, and satellite scale radii.

For each tested system, we select the semi-analytical evolution characterised by the $\alpha$ parameter that better reproduces the mass evolution of the satellite in N-body simulations. Such parameter regulates the rapidity of mass loss in our semi-analytical model, with higher values related to faster mass loss. We found a mild dependence of the best $\alpha$ with the initial orbital velocity, while no significant dependencies  with the satellite scale radius and orbital inclination are observed. Lower values of $\alpha$ were associated with more eccentric orbits, reflecting the need for a larger delay in mass loss due to faster pericenter passages.

Our model demonstrated excellent agreement with N-body simulations, accurately reproducing the satellite mass evolution, especially for systems with mildly and highly eccentric orbits. However, for systems with initial velocities close to $v_c$, a slight systematic overestimation of the satellite mass loss was observed.

This mass loss excess observed in systems with satellites on low-eccentricity orbits is likely influenced by two primary processes: tidal heating and satellite evaporation induced by mass truncation. To delve into this discrepancy, we run additional N-body simulations, where at each apocenter a re-orientation of star velocities along the tangential direction was performed. In systems where the satellite orbits within the galactic plane, the reorientation of star velocities mitigates the excess mass loss, aligning the simulation results with the predictions of our semi-analytical model. This suggests that, within the galactic plane, together with tidal stripping, satellite evaporation plays a dominant role in shaping the mass evolution. Still, outside the galactic plane, the reduction in excess mass loss is milder, and tidal heating effects become significant. This indicates that, in these configurations, both tidal heating and satellite evaporation contribute to the observed discrepancies between N-body simulations and the semi-analytical model.

Moreover, for orbits within the galactic plane, we compared our semi-analytical prescription for the satellite mass evolution with the instantaneous mass loss predicted using King's formula in reproducing the results of N-body simulations. We found that our model better reproduces the mass evolution in the simulations. It is important to stress that outside the galactic plane  - and in general in every non central potential- King's tidal radius is not well defined.

We then consider systems with both tidal stripping and dynamical friction effects. The semi-analytical model accurately reproduces both the orbital evolution and mass loss of the satellite.

These findings provide valuable insights into the complex interplay of tidal forces, dynamical friction, and the orbital parameters of satellite galaxies. Understanding these processes is crucial for accurately modelling the evolution of satellite galaxies within their host galactic environments.

\begin{acknowledgements}
We thank David Izquierdo-Villalba, Pedro Capelo, Lucio Mayer and Eugene Vasiliev for valuable discussions and suggestions.
MB acknowledges support provided by MUR under grant ``PNRR - Missione 4 Istruzione e Ricerca - Componente 2 Dalla Ricerca all'Impresa - Investimento 1.2 Finanziamento di progetti presentati da giovani ricercatori ID:SOE\_0163'' and by University of Milano-Bicocca under grant ``2022-NAZ-0482/B''. LV aknowledges support from MIUR under the grant PRIN 2017-MB8AEZ. AL acknowledges support by the PRIN MUR "2022935STW". EB acknowledges the financial support provided under the European Union's H2020 ERC Consolidator Grant ``Binary Massive Black Hole Astrophysics'' (B Massive, Grant Agreement: 818691).
EB acknowledges support from the European Union's Horizon Europe programme under the Marie Skłodowska-Curie grant agreement No 101105915 (TESIFA).

\end{acknowledgements}


\bibliographystyle{aa} % style aa.bst
\bibliography{biblio}

\begin{appendix} %First appendix
\section{Systems with orbits outside the galactic plane}
\begin{figure*}[]
%\centering
 \includegraphics[width=\hsize]{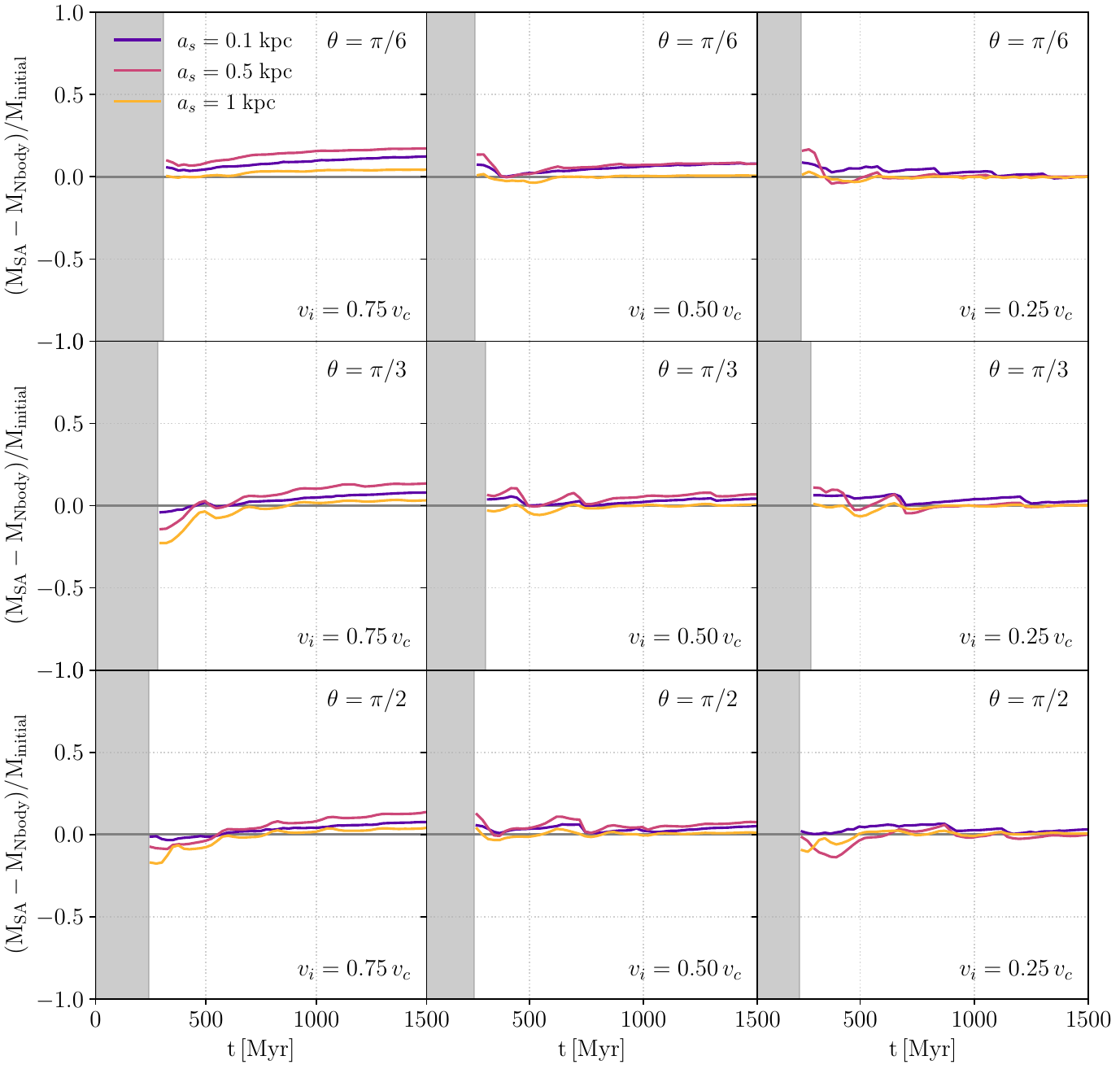}
\caption{Error in estimating the satellite mass for systems on inclined orbits with respect to the galactic plane and without dynamical friction. The line colors indicate different satellite scale radii. The columns represent different initial velocities of the satellite CoM, decreasing from left to right, while the rows illustrate varying orbital inclinations, increasing in angle from top to bottom. }
\label{Fig:Mass_err_OP_APO}%
\end{figure*}
In this section, we present the results for the systems with the satellite galaxy orbiting outside the galactic plane.
The columns represent different initial velocities of the satellite CoM, decreasing from left to right, while the rows illustrate varying orbital inclinations, increasing in angle from top to bottom.

\end{appendix}

\end{document}